# Quantized Dissipative Uncertain Model for Fractional T–S Fuzzy systems with Time-Varying Delays Under Networked Control System

**Muhammad Shamrooz Aslam[1], Hazrat Bilal[2,*], Summera Shamrooz[3]**

# ABSTARCT

This paper addressed with the quantized dissipative uncertain problem for *delayed fractional T–S Fuzzy system* for *event–triggered networked systems (E-NS)*, where the extended dissipativity analysis combines the $H_\infty$, dissipativity, $L_2 - L_\infty$ and passivity performance in a unified frame. To attain the high efficiency for available channel resources, measurement size decrease mechanism and *event–triggered scheme (ETS)* are proposed. **Firstly**, we present the *ETS* in which signal is transmitted through the channel with logical function then logarithmic quantization methodology is implemented for size reduction. **Then**, we transfer the original *delayed fractional T–S fuzzy systems* with the effect of quantization under *ETS* as induced communications delays. **Furthermore**, by employing the *associative Lyapunov functional* method in terms of linear matrix inequalities, adequate conditions for asymptotical stability is given. Moreover, we also construct the design fuzzy model for state–space filtering system. At last, a truck–trailer model is given to show the effectiveness of the proposed strategy.

# 1 Introduction

In the present era, the analysis of Singular time-delay systems has attained much attention due to its implication both in theoretical and practical applications (see e.g., [1-4] for a information). Singular time-delay systems [5], and [6], which are also called as descriptor time-delay systems are a type of hybrid model for dynamical systems consisting of a switching rule and a set of subsystems that coordinates the switching among them. The great benefit of such systems lies in that affluence of real-time systems or processes can be explained using switched models, such as networked control systems (NCSs), power systems, chemical processes, computer-controlled systems, and aircraft control systems. A huge number of results that related on switched systems have been investigated, for example, reliable filter based on switched systems suffered sensor failures is established in [7] asynchronously switched control systems is designed in [8]. In

1. Artificial Intelligence Research Institute, China University of Mining and Technology, Xuzhou, 2211106, China. Email address: shamroz_aslam@cumt.edu.cn (M.S Aslam)
2. College of Mechatronics and Control Engineering; and College of Computer Science and Software Engineering Shenzhen University, Shenzhen 518060, China. Email address: hbilal@mail.ustc.edu.cn (H. Bilal)
3. School of Mechanical and Electrical Engineering, China University of Mining and Technology, Xuzhou 221116, China. Email address: summera_shamroz@cumt.edu.cn (S. Shamroz) [9], the authors explore the properties of average dwell time for switched systems and switched systems has been reported with sampled–data control in [10-13]. It is worth mentioning that the above results do not consider the uncertain system. However, from a practical point of view, the industrial application and engineering system always have some uncertainties, which are induced by several factors, such as, parameter identification error, measurement error, approximate discretization and so on. Therefore, it is compulsory to examine the uncertain system when studying the singular time delay problem. Up until now, minute research results have taken into account the $H_\infty$ problem of uncertain systems, especially the uncertainty is included in delayed filter states.

Multi–agent systems (MAS) have emerged as a powerful framework for modeling and controlling large–scale, distributed, and intelligent systems where multiple autonomous agents interact to achieve individual or collective objectives [14, 15]. However, the intrinsic nonlinearities, uncertainties, and dynamic interactions among agents present significant challenges in achieving robust coordination and stability. To address these issues, the Takagi–Sugeno (T–S) fuzzy system has been widely adopted due to its ability to approximate complex nonlinear dynamics using a set of local linear models blended through fuzzy membership functions [16-18], and [19]. In the same way, the authors explored the properties of fuzzy membership function for various nonlinear models [20, 21]. By integrating T–S fuzzy modeling into MAS, researchers can design decentralized or distributed control strategies that handle agent heterogeneity and environmental disturbances with greater flexibility and precision. In particular, T–S fuzzy systems enable the construction of observer–based and state–feedback controllers that ensure consensus, formation, or cooperative tracking [22], and [23] and trajectory theory [24], and [25] even under switching topologies or communication delays. Recent advancements also explore stability analysis and performance guarantees using Lyapunov-based methods and Linear Matrix Inequalities (LMIs), making T-S fuzzy-based MAS control a promising direction for applications in robotics, autonomous vehicles, smart grids, and sensor networks.

The main object of this paper is to explore the properties of DC motor systems. DC motor system is the most general kind of motor system, which plays a significant role in the people□s daily life and production of industry. In the last few years, researchers have been extensively explored the problems of DC motor system in term of performance design and stability analysis [26-29]. Conversely, most of the research outcomes regard DC motor system treated as a linear system. Actually, in practical application, the DC motor system is, in fact, a complex plant that involves a diversity of nonlinearities, such as excitation nonlinearity, armature resistor nonlinearity, and saturation nonlinearity, etc. Therefore, from this point of view the practical and engineering application, the nonlinear DC motor system is preferred as the aim of this paper. To encounter the nonlinear problems of DC motor system, T-S fuzzy model with singular properties is implemented in this paper. One of the most important properties of the T-S fuzzy model is a universal approximation, in which we can smooth nonlinear function by a blending of some local linear system models. Due to this framework of the T-S fuzzy system, researchers have been a lot of explores on performance design and stability analysis and [30-32], and [33].

Please be aware that the aforementioned filters do not handle delays in real time. As a rule, the filter could have delays in its own operations; more specifically, in the case of the communication network channel, the plant could send a signal that the filter receives after the fact. Then, assuming controllers with delays in both states and inputs would be more important [46, 47], and [34]. Regrettably, the motivational problem of solitary T–S fuzzy delay systems remains unanswered and difficult, and has not yet been thoroughly investigated. Some of the ideas that drive our work come from here. Dead sensors, biassed sensors, and other types of failures (such as loss–of–efficiency, floating intensity, and stuck) have all been taken into account in power system analysis. Researching ways to detect when these sensor failures occur is increasingly relevant now. In contrast, random time delay is a common feature of signal transmission in many real–world systems; this feature can lead to filtering system instability or poor performance, which is why random time–delay systems have been a hot subject of study for decades. The use of *fuzzy LKFs*, which are derived from the fuzzy membership functions, is another aspect of our paper. These functions assist reduce the conservatism of design conditions and analysis for solitary T-S fuzzy systems with time delays since they provide additional information of the system model.

This paper addresses the problem of designing singular fuzzy filters with random time-varying delays for a class of continuous-time T-S fuzzy systems under the sensor fault. Main contributions and novelties are summarized as follows:

- Firstly, a general kind of fractional fuzzy controllers incorporating the input and state delays are established. Based on the delayed filters, the filtering error system is modeled as a singular fuzzy system with random time-varying delays.
- Secondly, the notion of $H_\infty$ performance is introduced, which capable to analyze the robustly asymptotically stability in a unified framework.
- Thirdly, single and double integrals are implemented to construct a *fuzzy LKF* involving fuzzy matrices, which produces less conservative delay-dependent conditions for the existence of the desired $H_\infty$ singular fuzzy filters. Finally, the DC motor example is presented to demonstrate the effectiveness of the proposed design methods.

*Notations* : The notations used in this paper are standard, and most of them are well used in the existing works. Specifically, for real symmetric matrices $X$ and $Y$, throughout this paper the notation $X \geq Y$ (respectively, $X > Y$) means that the $X - Y$ matrix is positive semi-definite (respectively, positive definite).. $|.|$ denotes the Euclidean norm for vectors and $\|.\|$ denotes the spectral norm for matrices.

# 2. System Description and Problem Formulation

In this paper, the general structure of the control problem is presented. In this research, the T–S fuzzy model is incorporated with nonlinear DC motor. The outputs of the DC motor are transmitted over the communication channel [35], and [36]. In what follows, we designed the DC motor with uncertainties under the T–S fuzzy system, sensor failure controller, filtering error system, respectively in a unified framework. In the published literature on DC motor system, mostly researcher focused on the armature resistor of DC motor system in which they considered armature resistor as a linear resistor [37], and [38]. However, the fact is contrary to it. Actually, in DC motor variation of armature resistor depends upon the voltage or current rating. So usually speaking, armature resistors are a nonlinear component in all DC motor systems.There are three main types of Nonlinear resistor:

- Current controlled nonlinear resistor. In this type, the terminal voltage is a single value function of its current for such nonlinear resistor.
- Voltage controlled nonlinear resistor. In this kind of current of such nonlinear resistor is a single value function of the voltage.
- Monotonic type nonlinear resistor. Characteristic of volt-ampere for such nonlinear resistor is monotonically decreasing or monotonically increasing, and at the same time, it is controlled by voltage and current and voltage.

Among the definitions of all the nonlinear resistors, the current controlled nonlinear resistor plays an important role in the nonlinear resistor, which is common in practical applications [39]. Therefore, it is more meaningful to assume armature resistor in a DC motor system as a current controlled nonlinear resistor.

Table 1: Notation used throughout the paper

| | |
|---|---|
| $\mathbf{L}_2[0,\infty)$ | space of square-integrable vector functions over $[0,\infty)$. |
| $diag(\dots)$ | Block-diagonal matrix. |
| $*$ | Used for the symmetry |
| T | Transpose of the matrix |
| I | Identity matrix of appropriate dimension |
| $1_n \in \mathbb{R}^n$ | vector of ones |
| $\mathbf{X} \geq \mathbf{Y}$ | $\mathbf{X}-\mathbf{Y}$ matrix is positive semi-definite |

In this article, we investigate the analysis and synthesis of nonlinear armature resistor with a DC motor system which consists of a nonlinear DC motor system [40]. The symbolic diagram of the nonlinear DC motor system is presented in this research, where a complex structure of the actual nonlinear DC motor system is converted into a simple schematic. Volt–ampere characteristic can be disclosed by function relation

$$u_R(t) = g(i(t)) \quad (1)$$

It's easy to plot the behavior of volt-ampere characteristic for the nonlinear resistor using the relation, which is given below

$$u_R(t) = 3i(t) + i^4(t) \quad (2)$$

the volt-ampere characteristic trajectory can be obtained, as exhibited in Figure 3. In the sense of a positive way of the armature circuit presented in Figure 2, by applying the *Kirchhoff□s Voltage Law (KVL)*, the electromotive force balance equation can be written as:

$$u_L(t) + e(t) + g(t) = u(t) \quad (3)$$

where $u_L(t)$ is the inductor voltage which is calculated from the current and voltage relation of the inductor $u_L(t) = L\frac{di}{dt}$; while $e(t)$ is the back electromotive force $e(t) = K_b\omega(t)$, and $K_b$ represents the electromotive force constant [41]. On the other side $e(t)$ is calculated by the armature winding cutting off the main flux when the armature rotates; applied voltage to nonlinear DC motor system is $u(t)$. Further going to the deep analysis of the nonlinear DC motor system, the torque balance equation:

$$\mathbb{T}(t) - \mathbb{T}_L(t) = \mathbb{J}\frac{d\omega(t)}{dt} \tag{4}$$

where $\mathbb{T}(t) = \mathbb{K}_a i(t)$ presents the electromagnetic torque, and $\mathbb{K}_a$ denotes the torque constant, in which $\mathbb{T}(t)$ express the rotation torque based on the rotor by the interaction between the rotor current and the magnetic flux under the rotating magnetic field; $\mathbb{T}_L(t) = \mathbb{B}\omega(t)$ is the required torque when the motor lifts the rotating load. It is known as load torque. In this paper, we assume an actual situation that there are the uncertainties, parameter perturbation and disturbance signal in the DC motor system. Then, by integrating (3) and (4), a dynamic model of DC motor system can be rewritten as follows:

$$\begin{cases} \mathbb{L}\dfrac{di(t)}{dt} = -g(i(t)) - (\mathbb{K}_b + \Delta_b(t))\omega(t) + u(t) \\ \mathbb{J}\dfrac{d\omega(t)}{dt} = \mathbb{K}_a i(t) - \mathbb{B}\omega(t) + \omega(t) \end{cases} \tag{5}$$

where $\Delta_b(t)$ is an uncertainty fulfilling $|\Delta_b(t)| < \mathbb{K}_b$; $\omega(t)$ presents the torque disturbance, which happened by the diversity of the practical working circumstances. By choosing $u(t)$ as the control input and rotation speed $\omega(t)$ and armature current $i(t)$ as the system states, then we can obtain the state space equation for the nonlinear DC motor system. In addition, since R is passive resistance, with $\omega(t) = 0$ the autonomous system is asymptotically stable [48], then one can choose $u(t) = 0$ for investigating the filtering problem.

## 2.1 Dynamics of Nonlinear DC Motor System

## 2.2 T-S Fuzzy Model

Let assume a nonlinear time-delay system approximated by a delayed T-S fuzzy singular model [49-51]:

**PlantRule** $i$:

**IF** $\{\varphi_1(t)$ is $\mathbb{W}_1^i\}$, $\{\varphi_2(t)$ is $\mathbb{W}_2^i\}$,...,$\{\varphi_p(t)$ is $\mathbb{W}_p^i\}$ **THEN**

$$
\begin{cases}
\mathbb{E}\dot{x}(t) & = & \bar{\mathbf{A}}_i(t)x(t)+\bar{\mathbf{A}}_{di}(t)x(t-d(t))+B_{\omega i}\omega(t) \\
z(t) & = & \bar{\mathbf{E}}_i(t)x(t)+\bar{\mathbf{E}}_{di}(t)x(t-d(t)) \\
y(t) & = & \mathbf{C}_i(t)x(t) \\
x(t) & = & \psi(t), \text{for each } t\in[-\bar{d},0]
\end{cases} \tag{6}
$$

where $x(t)\in\mathbf{R}_p$ presents the state vector; $z(t)\in\mathbf{R}_m$ denotes the output vector; $\mathbb{W}_j^i$, ($j=1,2,...,p$ , $i=1,2,...,r$) belongs to fuzzy sets, premise variable vectors are { $\varphi_1(t),\varphi_2(t),...,\varphi_p(t)$ }; $r$ is the symbol to compensate the number of IF-THEN rules; $y(t)\in\mathbf{R}_s$ presents the measure output vector; $\omega(t)\in\mathbf{R}_q$ belongs to the disturbance input vector which belongs to $\mathbf{L}_2[0,\infty)$. $\psi(t)$ shows the initial value function over the $t\in[-\bar{d},0]$. In this paper, the matrix $\mathbb{E}$ belongs to $\mathbf{R}_{p\times p}$ may be singular and supposed that rank $E=r\le p$ . Furthermore, $\bar{\mathbf{A}}_i(t)=\mathbf{A}_i+\Delta\mathbf{A}_i(t)$ , $\bar{\mathbf{A}}_{di}(t)=\mathbf{A}_{di}+\Delta\mathbf{A}_{di}(t)$ , $\bar{\mathbf{E}}_i(t)=\mathbf{E}_i+\Delta\mathbf{E}_i(t)$ and $\bar{\mathbf{E}}_{di}(t)=\mathbf{E}_{di}+\Delta\mathbf{E}_d(t)$, where $\mathbf{A}_i$, $\mathbf{A}_{di}$, $\mathbf{E}_i$ and $\mathbf{E}_{di}$ are known with proper dimension matrices. In the same way $\Delta\mathbf{A}_i(t)$ , $\Delta\mathbf{A}_{di}(t)$ , $\Delta\mathbf{E}_i(t)$ and $\Delta\mathbf{E}_{di}(t)$ are time-varying vectors over parametric uncertainties as given:

$$
\begin{bmatrix}\Delta\mathbf{A}_i(t) & \Delta\mathbf{A}_{di}(t) & \Delta\mathbf{E}_i(t) & \Delta\mathbf{E}_d(h)\end{bmatrix}=M_i\nabla(t)\begin{bmatrix}N_{1i} & N_{2i} & N_{3i} & N_{4i}\end{bmatrix} \tag{7}
$$

where $M_i$, $N_{ki}$ $k=1,2,3,4$ are the given constant matrices with proper dimension. Now we introduce the linear fractional form which was proposed in [52] including the norm bound uncertainties as a special case. $\nabla(t)$ presents the class of uncertainties which satisfy the conditions $\nabla(t)=\begin{bmatrix}I-G_i(t)L\end{bmatrix}^{-1}G_i(t)$ is said to be admissible, where $L$ is also given matrix satisfying the $I-LL^T>0$. On the other side, $G(t)$ is an unknown time dependent matrix with Lebesgue measurable bounded by $G_i(t)^TG_i(t)\le I$ [53] and [54]. Let $\lambda_i(\zeta(t))$ be the normalized membership function for inferred fuzzy sets $\alpha_i(\zeta(t))$. Then, by fuzzy blending, overall system can be written as:

$$
\begin{cases}
\mathbb{E}\dot{x}(t) & = & \sum_{i=1}^{r}\lambda_i(\zeta(t))\left\{\bar{\mathbf{A}}_i x(t)+\bar{\mathbf{A}}_{di}x(t-d(t))+B_i\omega(t)\right\} \\
z(t) & = & \sum_{i=1}^{r}\lambda_i(\zeta(t))\{\bar{\mathbf{E}}_i x(t)+\bar{\mathbf{E}}_{di}x(t-d(t))\} \\
y(t) & = & \sum_{i=1}^{r}\lambda_i(\zeta(t))\mathbf{C}_i x(t)
\end{cases} \tag{8}
$$

where $\lambda_i(\zeta(t))=\dfrac{\alpha_i(\zeta(t))}{\sum_{i=1}^{r}\alpha_i(\zeta(t))}$ with $\alpha_i(\zeta(t))=\prod_{j=1}^{q}\mathbb{W}_j^i(\zeta_j(t))$ in which $\mathbb{W}_j^i(\zeta_j(t))$ is the grade of the membership function of $\zeta_j(t)$. Also, we suppose $\alpha_i(\zeta(t)) \geq 0$, $i = 1, 2, ..., r$, and $\sum_{i=1}^{r} \alpha_i(\zeta(t)) > 0$, and $\lambda_i(\zeta(t))$ fulfill $\lambda_i(\zeta(t)) \geq 0$, $i = 1, 2, ..., r$, $\sum_{i=1}^{r} \lambda_i(\zeta(t)) = 1$ for any $\zeta(t)$.

Throughout the paper, we followed the following definition based on delayed T-S fuzzy singular system.

**Definition 1:** The delayed T-S fuzzy singular system

$$\begin{cases} \mathbb{E}\dot{x}(t) & = & \bar{\mathbf{A}}_i x(t) + \bar{\mathbf{A}}_{di} x(t - d(t)) \\ x(t) & = & \psi(t), \text{for each } t \in [-\bar{d}, 0] \end{cases} \tag{9}$$

is known as impulse free and regular, if the pair $(\mathbb{E}, \bar{\mathbf{A}}_i)$ is impulse free and regular.

Now we introduce the random time-delays for an above system which satisfy the follow assumptions:

**Assumption 1** *The time-varying delay $d(t)$ have upper and lower bound such that $d_m \leq d(t) \leq d_M$ and its probability distribution can be found i.e., let $d(t)$ takes values over $[d_m : d_0]$ or $(d_0, d_M]$ and $Prob\{d(t) \in [d_m : d_0]\} = \delta_0$ or $Prob\{d(t) \in (d_0 : d_M]\} = 1 - \delta_0$, where $d_m$, $d_0$ and $d_M$ are constant integers which satisfying $d_m \leq d_0 < d_M$, and $0 \leq \delta_0 \leq 1$. It is observed that in most practice e.g. power system, researchers take large values in terms of time delays but the probabilities of such values are very small. With this consideration, a scalar $d_0$ fulfils $d_m \leq d_0 \leq d_M$ is presented here.*

To present the probability distribution of $d(t)$ in terms of time delays, define following sets:

$$\mathfrak{D}_1 = \{t \mid d(t) \in [d_m : d_0]\}, and \mathfrak{D}_2 = \{t \mid d(t) \in (d_0 : d_M]\}, \tag{10}$$

In the same way, we define the mapping function as:

$$d_1(t) = \begin{cases} d(t), t \in \mathfrak{D}_1 \\ \bar{d}_m, t \in \mathfrak{D}_2 \end{cases} \tag{11}$$

$$d_2(t) = \begin{cases} d(t), t \in \mathfrak{D}_2 \\ \bar{d}_0, t \in \mathfrak{D}_1 \end{cases} \tag{12}$$

where $\bar{d}_m = [d_m, d_0]$ and $\bar{d}_0 = (d_0, d_M]$

**Assumption 2** *In this assumption, we define the conditions for time-varying delays of $d_1(t)$ and $d_2(t)$*

$$d_m \leq d_1(t) \leq d_0, \dot{d}_1(t) \leq \sigma_1$$
$$d_0 < d_2(t) \leq d_M, \dot{d}_2(t) \leq \sigma_2 \tag{13}$$

where $d_m$, $d_0$, $d_M$, $\sigma_1$ and $\sigma_2$ are prescribed positive constants. For more detail regarding probability distribution of time delays, researcher can refer the paper [55], and [56].

From (10), we conclude that $\mathfrak{D}_1 \cup \mathfrak{D}_2 = \mathfrak{Z}_{\geq 0}$, $\mathfrak{D}_1 \cap \mathfrak{D}_2 = \Theta$, where $\Theta$ is known as empty set. Its simple to check that $t \in \mathfrak{D}_1$ yields the event $d(t) \in [d_m, d_0]$ occurs, and $t \in \mathfrak{D}_2$ gives the event $d(t) \in (d_0, d_M]$ occurs.

Now we introduce the Bernoulli distributed stochastic variable:

$$\delta(t) = \begin{cases} 1, t \in \mathfrak{D}_1 \\ 0, t \in \mathfrak{D}_2 \end{cases} \tag{14}$$

With the definition of probability, we adopted two cases: [(a)]

1. $Prob\{\delta(t) = 1\} = Prob\{d(t) \in [d_m, d_0,]\} = \mathcal{L}[\delta(t)] = \delta_0$
2. $Prob\{\delta(t) = 0\} = Prob\{d(t) \in (d_0, d_M,]\} = 1 - \mathcal{L}[\delta(t)] = 1 - \delta_0$

Furthermore, its easy to observe that $\mathcal{L}[\delta(t) - \delta_0] = 0$ and $\mathcal{L}[(\delta(t) - \delta_0)^2] = \delta_0(1 - \delta_0)$. Now we incorporate the random delays with T-S fuzzy singular system (8). It can be rewritten as:

$$\begin{cases} \mathbb{E}\dot{\mathrm{x}}(t) & = & \sum_{i=1}^{r} \lambda_i(\zeta(t))\{\bar{\mathbf{A}}_i(t)\mathrm{x}(t) + \delta(t)\bar{\mathbf{A}}_{di}(t)\mathrm{x}(t - d_1(t)) + (1 - \delta(t))\bar{\mathbf{A}}_{di}(t)\mathrm{x}(t - d_2(t)) + B_i\omega(t)\} \\ \mathrm{z}(t) & = & \sum_{i=1}^{r} \lambda_i(\zeta(t))\{\bar{\mathbf{E}}_i(t)\mathrm{x}(t) + \bar{\mathbf{E}}_{di}(t)x(t - d(t))\} \\ \mathrm{y}(t) & = & \sum_{i=1}^{r} \lambda_i(\zeta(t))\mathbf{C}_i(t)\mathrm{x}(t) \end{cases} \tag{15}$$

## 2.3 Modeling of Sensor fault

In this section, we introduced the existing of sensor failure in the model which comes from [57-59], and [60].

$$\mathrm{y}^F(t) = \beta \mathrm{y}(t) \tag{16}$$

where $\mathrm{y}^F(t)$ is measured signal at the controller side from the sensor and $\beta = diag\{\beta_1, \beta_2, \beta_3, ..., \beta_s\}$ with $0 \le \underline{\beta}_\vartheta \le \beta_\vartheta \le \bar{\beta}_\vartheta \le 1$, $\vartheta = 1, 2, 3, ..., s$. For each $\underline{\beta}_\vartheta$, $\bar{\beta}_\vartheta$ and $\vartheta$ are given real constant, which have ability to admissible failures of $\vartheta^{th}$ sensor. When $\underline{\beta}_\vartheta = \bar{\beta}_\vartheta = 0$, its mean our system has gone in faulty state. Similarly, when $\underline{\beta}_\vartheta = \bar{\beta}_\vartheta = 1$ our system has no fault. In the case of partial failure $\vartheta^{th}$ sensor, the value of $\vartheta$ lies between $0 < \underline{\beta}_\vartheta < \bar{\beta}_\vartheta < 1$. Denote:

$$\begin{aligned} \hat{\vartheta} &= diag\left[\frac{\underline{\beta}_1 + \bar{\beta}_1}{2}, \frac{\underline{\beta}_2 + \bar{\beta}_2}{2}, ..., \frac{\underline{\beta}_s + \bar{\beta}_s}{2}\right] \\ \breve{\vartheta} &= diag\left[\frac{\underline{\beta}_1 - \bar{\beta}_1}{2}, \frac{\underline{\beta}_2 - \bar{\beta}_2}{2}, ..., \frac{\underline{\beta}_s - \bar{\beta}_s}{2}\right] \end{aligned} \tag{17}$$

Furthermore, we can write the $\beta$ as

$$\beta = \hat{\vartheta} + diag\{\rho_1, \rho_2, \rho_3, ..., \rho_s\} \tag{18}$$

where $\rho_\vartheta \leq \frac{\underline{\beta}_\vartheta - \overline{\beta}_\vartheta}{2}$, $\vartheta = 1, 2, 3, ..., s$.

Now, we consider a delayed controller For the T-S fuzzy singular system described by (8) in the following form:

$$\begin{cases} \mathbb{E}_f \dot{\mathrm{x}}_f(t) &= \bar{\mathbf{A}}_{fi}\mathrm{x}_f(t) + \bar{\mathbf{A}}_{\tau fi}\mathrm{x}_f(t-\tau(t)) + \mathbf{B}_{fi}\mathrm{y}^F(t) \\ \mathrm{z}_f(t) &= \bar{\mathbf{E}}_{fi}\mathrm{x}_f(t) + \bar{\mathbf{E}}_{\tau fi}\mathrm{x}_f(t-\tau(t)) + \mathbf{D}_{fi}\mathrm{y}^F(t) \\ \mathrm{x}_f(t) &= \psi_f(t), \quad t \in [\overline{\tau}, 0] \end{cases} \tag{19}$$

where $\mathrm{x}_f(t) \in \mathbf{R}_p$ and $\mathrm{z}_f(t) \in \mathbf{R}_m$ denote the input sate and output vector of the controller respectively; $\phi_f(t)$ is the initial condition; $\tau(t)$ presents the delay appeared in the controller state, which satisfy the condition $0 \leq \tau(t) \leq \overline{\tau},\ \dot{\tau}(t) \leq \sigma_3$ ; the matrices $\bar{\mathbf{A}}_{fi} = \mathbf{A}_{fi} + \Delta\mathbf{A}_{fi}(t)$ , $\bar{\mathbf{A}}_{\tau fi} = \mathbf{A}_{\tau fi} + \Delta\mathbf{A}_{\tau fi}(t)$ , $\bar{\mathbf{E}}_{fi} = \mathbf{E}_{fi} + \Delta\mathbf{E}_{fi}(t)$ and $\bar{\mathbf{E}}_{\tau fi} = \mathbf{E}_{\tau fi} + \Delta\mathbf{E}_{\tau fi}(t)$ , where $\mathbf{A}_{fi}$ , $\mathbf{A}_{\tau fi}$ , $\mathbf{E}_{fi}$ and $\mathbf{E}_{\tau fi}$ are the controller parameter to be determined later with fulfil the following condition. In the same way $\Delta\mathbf{A}_i(t)$ , $\Delta\mathbf{A}_{\tau i}(t)$ , $\Delta\mathbf{E}_i(t)$ and $\Delta\mathbf{E}_{\tau i}(t)$ are time-varying vectors over parametric uncertainties as given:

$$\begin{bmatrix} \Delta\mathbf{A}_{fi}(t) & \Delta\mathbf{A}_{\tau fi}(t) & \Delta\mathbf{E}_{fi}(t) & \Delta\mathbf{E}_{\tau fi}(t) \\ \end{bmatrix} = M_i \Upsilon_i(t) \begin{bmatrix} N_{5i} & N_{6i} & N_{7i} & N_{8i} \end{bmatrix} \tag{20}$$

where $\Upsilon_i(t)$ has the property of time-varying for unknown matrix function with Lebesgue norm measurable elements fulfil the condition $\Upsilon_i(t)^T \Upsilon_i(t) \leq I$ and $M_i$ , $N_{li}$ , $l = 5, 6, 7, 8$ are the given constant matrices with proper dimension. It will be well known that these parametric matrices have the characteristics of nonlinearity with respect to $\lambda_i(\zeta)$ . Thus, the controller in the form of (8) is actually non–PDC, which is more general than the parallel distributed compensation (PDC) [42], and [43].

In above–mentioned system, property of time–delay is supposed to be random and fulfill the following assumption Calculate the filtering error as $\nu(t) = \mathrm{z}(t) - \mathrm{z}_f(t)$ and the augmented vector $\zeta(t) = \begin{bmatrix} \mathrm{x}(t)^T & \mathrm{x}_f(t)^T \end{bmatrix}^T$ . Then, combining the system (8) and the controller (19) yields to the filtering error system:

$$\begin{cases} \overline{\mathbb{E}}\dot{\zeta}(t) = \mathcal{A}\zeta(t) + \sum_{i=1}^{2} \mathcal{A}_{d_i}\zeta(t - d_i(t)) + \mathcal{A}_\tau\zeta(t - \tau(t)) + \mathcal{B}\omega(t) \\ \nu(t) = \mathcal{E}\zeta(t) + \sum_{i=1}^{2} \mathcal{E}_{d_i}\zeta(t - d_i(t)) + \mathcal{E}_\tau\zeta(t - \tau(t)) \end{cases} \tag{21}$$

where

$$\mathcal{A}=\begin{bmatrix}\bar{\mathbf{A}}_i & 0\\ \mathbf{B}_{fi}\beta\mathbf{C}_i & \bar{\mathbf{A}}_{fi}\end{bmatrix},\mathcal{A}_{d_1}=\begin{bmatrix}\delta(t)\bar{\mathbf{A}}_{di} & 0\\ 0 & 0\end{bmatrix},\mathcal{A}_{d_2}=\begin{bmatrix}(1-\delta(t))\bar{\mathbf{A}}_{di} & 0\\ 0 & 0\end{bmatrix},$$

$$\mathcal{A}_\tau=\begin{bmatrix}0 & 0\\ 0 & \bar{\mathbf{A}}_{\tau fi}\end{bmatrix},\mathcal{B}=\begin{bmatrix}B_i\\ 0\end{bmatrix},\bar{\mathbb{E}}=diag\{\mathbb{E},\mathbb{E}_f\},\mathcal{E}=\begin{bmatrix}\bar{\mathbf{E}}_i+\mathbf{D}_{fi}\beta\mathbf{C}_i & -\bar{\mathbf{E}}_{fi}\end{bmatrix}$$

$$\mathcal{E}_{d_1}=\begin{bmatrix}\delta(t)\bar{\mathbf{E}}_{di} & 0\end{bmatrix},\mathcal{E}_{d_2}=\begin{bmatrix}(1-\delta(t))\bar{\mathbf{E}}_{di} & 0\end{bmatrix},\mathcal{E}_\tau=\begin{bmatrix}0 & -\bar{\mathbf{E}}_{\tau fi}\end{bmatrix}$$

The main objective of this article is to attain delay–dependent conditions for the existence of resilient delayed T–S fuzzy singular controller (21) assuring that the disturbance $\omega(t)$ and the filtering error $\nu(t)$ satisfies the $H_\infty$ performance constraint. Therefore, we quote the concept of the $H_\infty$ performance index as follows.

**Definition 2** *: Uncertain T-S fuzzy singular system (21) is said to be robust stable with prescribed disturbance level* $\gamma$ *for all* $L_2\in[0,\infty]$ *under the response* $\nu(t)$ *with zero initial conditions, i.e.,* $\psi(t)=0$*:*

$$\mathbb{E}\{\int_0^\infty \nu(t)^T\nu(t)\}\le\gamma^2\int_0^\infty\omega^T(t)\omega(t)dt$$

# 3 Main Results

In this section, we are going to present a delay–dependent conditions for designing the delayed singular fuzzy controller in the form of (8) such that the filtering error system (21) is asymptomatically stable, which comes from [44], and [45]. First, we present the following theorem, which give general conditions for the system (21).

**Theorem 1** *For the give scalars* $\gamma>0$*, filtering error system (21) is impulse-free with the* $H_\infty$ *performance index for any time-varying delays* $\tau_k(t)$ *satisfying Assumption 2, if there exist matrices* $\mathbf{S}_k>0$*,* $\mathbf{W}_k>0$*,* $\mathbf{P}_i>0$*,* $\mathbf{Q}_{ki}>0$*,* $\mathbf{Q}_4>0$ $\mathbf{R}_{ki}>0$*,* $\mathbf{Z}_{ki}>0$*,* $\mathbf{M}_{ki}$*, such that*

$$\Gamma_{1i}:=\dot{\mathbf{Q}}_{ki}+\dot{\mathbf{R}}_{ki}-\mathbf{S}_k<0,\quad \Gamma_{2i}:=\dot{\mathbf{R}}_{ki}-\mathbf{S}_k<0,\quad \Gamma_{3i}:=\dot{\mathbf{Z}}_{ki}-\bar{a}_k^{-1}\mathbf{W}_k<0,\ \textit{and}\ \begin{bmatrix}\mathbf{Z}_{ki} & \mathbf{M}_{ki}\\ * & \mathbf{Z}_{ki}\end{bmatrix}>0,$$

$k=1,2,3$*, and the following inequalities hold:*

$$\bar{\mathbb{E}}\mathbf{P}_i=\mathbf{P}_i^T\bar{\mathbb{E}}\ge 0 \tag{22}$$

$$\begin{bmatrix}\Gamma_{ij} & \mathbf{P}_i\mathcal{B} & \mathbb{A}_i^T\mathbf{P}_i & \mathfrak{E}_i^T\\ * & -\gamma^2 I & \mathcal{B}_i^T\mathbf{P}_i & 0\\ * & * & \hat{\mathbb{Z}}_i-2\mathbf{P}_i & 0\\ * & * & * & -I\end{bmatrix}<0 \tag{23}$$

where $\hat{\mathbb{Z}}_i = \sum_{k=1}^{3}(\bar{a}_k^2 \mathbf{Z}_{ki} + 1/2\bar{a}_k^2 \mathbf{W}_k)$, and

$$\Gamma_{ij} = \begin{bmatrix} \Gamma_{ij}^{11} & \Gamma_{ij}^{12} & \Gamma_{ij}^{13} & \Gamma_{ij}^{14} & \Gamma_{ij}^{15} & \Gamma_{ij}^{16} & \Gamma_{ij}^{17} & 0 \\ * & \Gamma_{ij}^{21} & 0 & \varphi_{ij}^{21} & 0 & 0 & 0 & 0 \\ * & * & \Gamma_{ij}^{22} & 0 & \varphi_{ij}^{22} & 0 & 0 & 0 \\ * & * & * & \psi_{ij}^{21} & 0 & 0 & 0 & 0 \\ * & * & * & * & \psi_{ij}^{22} & 0 & 0 & 0 \\ * & * & * & * & * & \Gamma_{ij}^{23} & \varphi_{ij}^{23} & 0 \\ * & * & * & * & * & * & \psi_{ij}^{23} & 0 \\ * & * & * & * & * & * & * & -\mathbf{Q}_4 \end{bmatrix}$$

$$\mathbb{A}_i = \begin{bmatrix} \mathcal{A} & \mathcal{A}_{d_1} & \mathcal{A}_{d_2} & 0 & 0 & \mathcal{A}_\tau & 0 & 0 \end{bmatrix}, \quad \mathfrak{E}_i = \begin{bmatrix} \mathcal{E} & \mathcal{E}_{d_1} & \mathcal{E}_{d_2} & 0 & 0 & \mathcal{E}_\tau & 0 & 0 \end{bmatrix}$$

$$\Gamma_{ij}^{11} = \dot{\mathbf{P}}_i + \mathbf{P}_i\mathcal{A} + \mathcal{A}^T\mathbf{P}_i + \sum_{l=1}^{3}\{\mathbf{Q}_{ki} + \mathbf{R}_{ki} + \bar{a}_k\mathbf{S}_k - \bar{\mathbb{E}}^T\mathbf{Z}_{ki}\bar{\mathbb{E}}\} + \mathbf{Q}_4$$

$$\Gamma_{ij}^{12} = \mathbf{P}_i\mathcal{A}_{d_1} + \varphi_{ij}^{21}, \Gamma_{ij}^{13} = \mathbf{P}_i\mathcal{A}_{d_2} + \varphi_{ij}^{22}, \Gamma_{ij}^{14} = \bar{\mathbb{E}}^T\mathbf{M}_{1i}\bar{\mathbb{E}}$$

$$\Gamma_{ij}^{15} = \bar{\mathbb{E}}^T\mathbf{M}_{2i}\bar{\mathbb{E}}, \Gamma_{ij}^{16} = \mathbf{P}_i\mathcal{A}_\tau + \varphi_{ij}^{23}, \Gamma_{ij}^{17} = \bar{\mathbb{E}}^T\mathbf{M}_{3i}\bar{\mathbb{E}}$$

$$\Gamma_{ij}^{2\ell} = -(1-\sigma_\ell)\mathbf{Q}_{\ell i} - 2\bar{\mathbb{E}}^T\mathbf{Z}_{\ell i}\bar{\mathbb{E}} + \bar{\mathbb{E}}^T\mathbf{M}_{\ell i}\bar{\mathbb{E}} + \bar{\mathbb{E}}^T\mathbf{M}_{\ell i}^T\bar{\mathbb{E}}, \quad \psi_{ij}^{2\ell} = -\bar{\mathbb{E}}^T\mathbf{Z}_{\ell i}\bar{\mathbb{E}} - \bar{\mathbb{E}}^T\mathbf{R}_{\ell i}\bar{\mathbb{E}}$$

$$\varphi_{ij}^{2\ell} = \bar{\mathbb{E}}^T\mathbf{Z}_{\ell i}\bar{\mathbb{E}} - \bar{\mathbb{E}}^T\mathbf{M}_{\ell i}\bar{\mathbb{E}}, \quad \ell = 1,2,3$$

Based upon the above-mentioned conditions in Theorem 1, we attain the following theorem, which presents general conditions for the existence of desired filters.

**Assumption 3** *There exist real constant scalars* $\lambda_i$ *such that* $\dot{\lambda}_i \le \rho_i, i = 1,2,3,\cdots,r$.

Under this assumption, we obtain LMI-based conditions for the existence of the desired filters, which are given in the following theorem.

**Remark 1** *General conditions are given in Theorem 1 for the existence of desired filters. When the LMI conditions in the theorem have feasible solutions, then controller parameters can be obtained according to (30). Although, the LMI conditions in Theorem 1 are based on the membership functions, which are usually complicated to be solved. Thus, it is compulsory to convert into the conditions in Theorem 1 to strict LMIs. To this end, we have to make an assumption on the rate of function* $h$.

**Assumption 4** *A real constant scalars exist* $\rho_i$ *such that* $\dot{h}_i \le \rho_i, i = 1,2,3,\cdots,r$.

**Theorem 2** *For the give scalars* $\gamma > 0$, *filtering error system (21) is impulse-free with the* $H_\infty$ *performance index for any time-varying delays* $\tau_k(t)$ *satisfying Assumption 2, if there exist* *matrices* $\breve{\mathbf{S}}_k > 0$, $\breve{\mathbf{W}}_k > 0$, $\mathbf{X}_i$, $\mathbf{Y}_j$, $\breve{\mathbf{Q}}_{ki} > 0$, $\breve{\mathbf{Q}}_4 > 0$ $\breve{\mathbf{R}}_{ki} > 0$, $\breve{\mathbf{Z}}_{ki} > 0$, $\breve{\mathbf{M}}_{ki}$, $\mathbf{L}_k$, $\mathbf{K}_k$, $\mathbf{N}_k$, $\breve{\mathbf{A}}_{fi}$, $\breve{\mathbf{A}}_{\tau fi}$, $\breve{\mathbf{B}}_{fi}$, $\breve{\mathbf{E}}_{fi}$, $\breve{\mathbf{E}}_{\tau fi}$ *and* $\breve{\mathbf{D}}_{fi}$ $k = 1,2,3$ *and* $i = 1,2,\cdots,r$ *such that the following inequalities hold for* $k = 1,2,3$ *and* $i, j = 1,2,\cdots,r$*:*

$$\bar{\mathbb{E}}^T \mathbf{X}_i = \mathbf{X}_i^T \bar{\mathbb{E}} \geq 0 \tag{24}$$

$$\bar{\mathbb{E}}^T \mathbf{Y}_j = \mathbf{Y}_j^T \bar{\mathbb{E}} \geq 0 \tag{25}$$

$$\bar{\mathbb{E}}^T (\mathbf{X}_i - \mathbf{Y}_j) \geq 0 \tag{26}$$

$$\breve{\Gamma}_k = \begin{cases} \breve{\Gamma}_{A_k} < 0 \\ \breve{\Gamma}_{B_k} > 0, \end{cases} \quad k = 1,2,3 \tag{27}$$

$$(\mathbf{X}_i + \mathbb{X}_0) > 0, \quad \begin{bmatrix} \breve{\mathbf{Z}}_{ki} & \breve{\mathbf{M}}_{ki} \\ * & \breve{\mathbf{Z}}_{ki} \end{bmatrix} > 0, \tag{28}$$

$$\Xi_{ij} + \Xi_{ji} < 0, i < j \tag{29}$$

where

$$\Xi_{ij} = \begin{bmatrix} \breve{\Gamma}_{ij} & \Theta_{5ij} & \Xi_{ij}^{a^T} & \Xi_{ij}^{b^T} \\ * & -\gamma^2 I & \Theta_{5ij}^T & 0 \\ * & * & \mathbb{Z}_i - 2\vartheta_i & 0 \\ * & * & * & -I \end{bmatrix}$$

$$\breve{\Gamma}_{ij} = \begin{bmatrix} \breve{\Gamma}_{ij}^{11} & \breve{\Gamma}_{ij}^{12} & \breve{\Gamma}_{ij}^{13} & \breve{\Gamma}_{ij}^{14} & \breve{\Gamma}_{ij}^{15} & \breve{\Gamma}_{ij}^{16} & \breve{\Gamma}_{ij}^{17} & 0 \\ * & \breve{\Gamma}_{ij}^{21} & 0 & \breve{\varphi}_{ij}^{21} & 0 & 0 & 0 & 0 \\ * & * & \breve{\Gamma}_{ij}^{22} & 0 & \breve{\varphi}_{ij}^{22} & 0 & 0 & 0 \\ * & * & * & \breve{\psi}_{ij}^{21} & 0 & 0 & 0 & 0 \\ * & * & * & * & \breve{\psi}_{ij}^{22} & 0 & 0 & 0 \\ * & * & * & * & * & \breve{\Gamma}_{ij}^{23} & \breve{\varphi}_{ij}^{23} & 0 \\ * & * & * & * & * & * & \breve{\psi}_{ij}^{23} & 0 \\ * & * & * & * & * & * & * & -\breve{\mathbf{Q}}_4 \end{bmatrix}$$

$$\breve{\Gamma}_{ij}^{11} = \sum_{i=1}^{r} \{ \begin{bmatrix} I \\ 0 \end{bmatrix} \rho_i (\mathbf{X}_i + \mathbb{X}_0) \begin{bmatrix} I & 0 \end{bmatrix} \} + \Theta_{1ij} + \Theta_{1ij}^T + \sum_{l=1}^{3} \{ \breve{\mathbf{Q}}_{ki} + \breve{\mathbf{R}}_{ki} + \bar{a}_k \breve{\mathbf{S}}_k - \bar{\mathbb{E}}^T \breve{\mathbf{Z}}_{ki} \bar{\mathbb{E}} \} + \breve{\mathbf{Q}}_4$$

$$\breve{\Gamma}_{ij}^{12} = \Theta_{2ij} + \breve{\varphi}_{ij}^{21}, \quad \breve{\Gamma}_{ij}^{13} = \Theta_{3ij} + \breve{\varphi}_{ij}^{22}, \quad \breve{\Gamma}_{ij}^{14} = \bar{\mathbb{E}}^T \breve{\mathbf{M}}_{1i} \bar{\mathbb{E}}$$

$$\breve{\Gamma}_{ij}^{15} = \bar{\mathbb{E}}^T \breve{\mathbf{M}}_{2i} \bar{\mathbb{E}}, \quad \breve{\Gamma}_{ij}^{16} = \Theta_{4ij} + \breve{\varphi}_{ij}^{23}, \breve{\Gamma}_{ij}^{17} = \bar{\mathbb{E}}^T \breve{\mathbf{M}}_{3i} \bar{\mathbb{E}}$$

$$\breve{\Gamma}_{ij}^{2\ell} = -(1-\sigma_\ell) \breve{\mathbf{Q}}_{\ell i} - 2\bar{\mathbb{E}}^T \breve{\mathbf{Z}}_{\ell i} \bar{\mathbb{E}} + \bar{\mathbb{E}}^T \breve{\mathbf{M}}_{\ell i} \bar{\mathbb{E}} + \bar{\mathbb{E}}^T \breve{\mathbf{M}}_{\ell i}^T \bar{\mathbb{E}}, \quad \breve{\psi}_{ij}^{2\ell} = -\bar{\mathbb{E}}^T \breve{\mathbf{Z}}_{\ell i} \bar{\mathbb{E}} - \bar{\mathbb{E}}^T \breve{\mathbf{R}}_{\ell i} \bar{\mathbb{E}}$$

$$\breve{\varphi}_{ij}^{2\ell} = \bar{\mathbb{E}}^T \breve{\mathbf{Z}}_{\ell i} \bar{\mathbb{E}} - \bar{\mathbb{E}}^T \breve{\mathbf{M}}_{\ell i} \bar{\mathbb{E}}, \quad \ell = 1,2,3$$

$$(\breve{\Gamma}_{A_1},\breve{\Gamma}_{B_1}) = (\sum_{i=1}^{r}\rho_i(\breve{\mathbf{Q}}_{ki}+\breve{\mathbf{R}}_{ki}+\mathbf{L}_k)-\breve{\mathbf{S}}_k,\breve{\mathbf{Q}}_{ki}+\breve{\mathbf{R}}_{ki}+\mathbf{L}_k)$$

$$(\breve{\Gamma}_{A_2},\breve{\Gamma}_{B_2}) = (\sum_{i=1}^{r}\rho_i(\breve{\mathbf{R}}_{ki}+\mathbf{K}_k)-\breve{\mathbf{S}}_k,\breve{\mathbf{R}}_{ki}+\mathbf{K}_k)$$

$$(\breve{\Gamma}_{A_3},\breve{\Gamma}_{B_3}) = (\sum_{i=1}^{r}\rho_i(\breve{\mathbf{Z}}_{ki}+\mathbf{N}_k)-\frac{1}{\bar{a}_k}\breve{\mathbf{W}}_k,\breve{\mathbf{Z}}_{ki}+\mathbf{N}_k$$

$$\Xi_{ij}^{a} = \begin{bmatrix}\Theta_{1ij} & \Theta_{2ij} & \Theta_{3ij} & 0 & 0 & \Theta_{4ij} & 0 & 0\end{bmatrix},\quad \Xi_{ij}^{b} = \begin{bmatrix}\mathcal{F}_{1ij} & \mathcal{F}_{2ij} & \mathcal{F}_{3ij} & 0 & 0 & \mathcal{F}_{4ij} & 0 & 0\end{bmatrix}$$

$$\Theta_{1ij} = \begin{bmatrix}\mathbf{Y}_j^T\bar{\mathbf{A}}_i & \mathbf{Y}_j^T\bar{\mathbf{A}}_i \\ \mathbf{X}_i^T\bar{\mathbf{A}}_i+\breve{\mathbf{B}}_{fj}\beta\mathbf{C}_i+\breve{\mathbf{A}}_{fj} & \mathbf{X}_i^T\bar{\mathbf{A}}_i+\breve{\mathbf{B}}_{fj}\beta\mathbf{C}_i \\ & \end{bmatrix},\quad \Theta_{2ij} = \delta_0\begin{bmatrix}\mathbf{Y}_j^T\bar{\mathbf{A}}_{di} & \mathbf{Y}_j^T\bar{\mathbf{A}}_{di} \\ \mathbf{X}_i^T\bar{\mathbf{A}}_{di} & \mathbf{X}_i^T\bar{\mathbf{A}}_{di} \\ & \end{bmatrix}$$

$$\Theta_{3ij} = (1-\delta_0)\Theta_{2ij}\quad \Theta_{4ij} = \begin{bmatrix}0 & 0 \\ \breve{\mathbf{A}}_{\tau fj} & 0 \\ & \end{bmatrix},\quad \Theta_{5ij} = \begin{bmatrix}\mathbf{Y}_j^T B_i \\ \mathbf{X}_i^T B_i \\ \end{bmatrix},\quad \vartheta_{ij} = \begin{bmatrix}\mathbf{Y}_j & \mathbf{Y}_j \\ \mathbf{Y}_j & \mathbf{X}_i \\ & \end{bmatrix}$$

$$\mathcal{F}_{1ij} = \begin{bmatrix}\bar{\mathbf{E}}_i+\breve{\mathbf{D}}_{fj}\beta\mathbf{C}_i-\breve{\mathbf{E}}_{fj} & \bar{\mathbf{E}}_i+\breve{\mathbf{D}}_{fj}\beta\mathbf{C}_i \\ & \end{bmatrix},\qquad \mathcal{F}_{2ij} = \delta_0\begin{bmatrix}\bar{\mathbf{E}}_{di} & \bar{\mathbf{E}}_{di} \\ & \end{bmatrix},\mathcal{F}_{3ij} = (1-\delta_0)\mathcal{F}_{2ij}$$

$$\mathcal{F}_{4ij} = \begin{bmatrix}-\breve{\mathbf{E}}_{\tau fi} & 0 \\ & \end{bmatrix},\quad \mathbb{Z}_i = \sum_{k=1}^{3}(\bar{a}_k^2\breve{\mathbf{Z}}_{ki}+1/2\bar{a}_k^2\breve{\mathbf{W}}_k),$$

In the above discussion, there exist nonsingular matrices $U$, $\bar{U}$, $W$ and $\bar{W}$ such that

$$\bar{\mathbb{E}}^T\bar{U} = U^T\bar{\mathbb{E}},\qquad \bar{\mathbb{E}}W = \bar{W}^T\bar{\mathbb{E}}^T$$

$$\mathbf{X}_i\mathbf{Y}_j^{-1} = I-\bar{U}W,\qquad \mathbf{Y}_j^{-1}\mathbf{X}_i = I-\bar{W}U$$

In this case, the parameters of the desired controller (19) can be computed according to the following equalities:

$$\begin{cases}\breve{\mathbf{A}}_{fj} = U^T\bar{\mathbf{A}}_{fi}W\mathbf{Y}_j, \\ \breve{\mathbf{E}}_{fj} = \bar{\mathbf{E}}_{fj}W\mathbf{Y}_j, \\ \breve{\mathbf{B}}_{fj} = U^T\bar{\mathbf{B}}_{fj},\end{cases}\qquad \begin{cases}\breve{\mathbf{A}}_{\tau fj} = U^T\bar{\mathbf{A}}_{\tau fj}W\mathbf{Y}_j, \\ \breve{\mathbf{E}}_{\tau fj} = \bar{\mathbf{E}}_{\tau fj}W\mathbf{Y}_j, \\ \breve{\mathbf{D}}_{fj} = \bar{\mathbf{D}}_{fj}\end{cases}\tag{30}$$

**Theorem 3** *For the give scalars* $\gamma > 0$*, filtering error system (21) is impulse-free with the* $H_\infty$ *performance index for any time-varying delays* $\tau_k(t)$ *satisfying Assumption 2, if there exist matrices* $\breve{\mathbf{S}}_k > 0$, $\breve{\mathbf{W}}_k > 0$, $\mathbf{X}_i$, $\mathbf{Y}_j$, $\breve{\mathbf{Q}}_{ki} > 0$, $\breve{\mathbf{Q}}_4 > 0$ $\breve{\mathbf{R}}_{ki} > 0$, $\breve{\mathbf{Z}}_{ki} > 0$, $\breve{\mathbf{M}}_{ki}$, $\mathbf{L}_k$, $\mathbf{K}_k$, $\mathbf{N}_k$, $\breve{\mathbf{A}}_{fi}$, $\breve{\mathbf{A}}_{\tau fi}$, $\breve{\mathbf{B}}_{fi}$, $\breve{\mathbf{E}}_{fi}$, $\breve{\mathbf{E}}_{\tau fi}$ *and* $\breve{\mathbf{D}}_{fi}$ $k = 1,2,3$ $i = 1,2,\cdots,r$ *and the scaler* $\varepsilon_1$ *and* $\varepsilon_2$ *such that the following inequalities hold for* $k = 1,2,3$ *and* $i,j = 1,2,\cdots,r$ *:*

$$\prod_{ij} = \begin{bmatrix} \Xi_{ij} & \prod_{ij}^{1} & \prod_{ij}^{2} \\ * & \varepsilon_1 & 0 \\ * & * & \varepsilon_2 \end{bmatrix} < 0, \tag{31}$$

where

$$\prod_{ij}^{1} = \begin{bmatrix} \prod_{ij}^{1} & \prod_{ij}^{2} \end{bmatrix}, \prod_{ij}^{2} = \begin{bmatrix} \prod_{ij}^{3} & \prod_{ij}^{4} \end{bmatrix}, \quad \varepsilon_f = diag\{-\varepsilon_f I, -\varepsilon_f^{-1} I\} \quad f = 1,2$$

$$\prod_{ij}^{1} = \mathrm{col}\begin{bmatrix} \prod_{ij}^{11} & \delta_0 \prod_{ij}^{11} & (1-\delta_0)\prod_{ij}^{11} & 0 & 0 & 0 & 0 & 0 \end{bmatrix}, \quad \prod_{ij}^{2} = \mathrm{col}\begin{bmatrix} \prod_{ij}^{21} & \prod_{ij}^{22} & \prod_{ij}^{23} & 0 & 0 & 0 & 0 & 0 \end{bmatrix}$$

$$\prod_{ij}^{11} = \begin{bmatrix} \mathbf{Y}_j^T M_i \\ \mathbf{X}_i^T M_i \end{bmatrix}, \quad \prod_{ij}^{2\ell} = \begin{bmatrix} N_{\ell i}^T \\ N_{\ell i}^T \end{bmatrix}, \qquad \ell = 1,2,3$$

$$\prod_{ij}^{3} = \mathrm{col}\begin{bmatrix} 1_2 \otimes M_i & -\delta_0 M_i & (1-\delta_0)M_i & 0 & 0 & 1_2 \otimes M_i & 0 & 0 \end{bmatrix}$$

$$\prod_{ij}^{4} = \mathrm{col}\begin{bmatrix} \begin{bmatrix} N_{7i}^T \\ N_{5i}^T \end{bmatrix} & 1_2 \otimes N_{4i} & \begin{bmatrix} N_{4i}^T \\ 0 \end{bmatrix} & 0 & 0 & \begin{bmatrix} N_{8i}^T \\ N_{5i}^T \end{bmatrix} & 0 & 0 \end{bmatrix}$$

Remaing parameters are same in mentioned in Theorem 2.

# 4 Numerical Example

In the proceeding examples, we elaborated the effectiveness of the proposed algorithm for delayed singular fuzzy controller design in this paper.

## 4.1 Example 1

In this example, we considered the singular T–S fuzzy system (8) with time-varying delays investigated in [61-66] with following parameters:

$$\mathbb{E} = \begin{bmatrix} 1 & 0 \\ 0 & 1 \end{bmatrix}$$

The parameters of Fuzzy matrices are:

$$\begin{bmatrix} \mathbf{A}_1 & \mathbf{A}_{d1} & B_{\omega 1} & \mathbf{C}_1' & \mathbf{E}_1' \\ \mathbb{A}_2 & \mathbf{A}_{d2} & B_{\omega 2} & \mathbf{C}_2' & \mathbf{E}_2' \end{bmatrix} = \begin{bmatrix} -2.1 & 0.1 & -1.1 & 0.1 & 1 & 1 & 1 \\ 1 & -2 & -0.8 & -0.9 & -0.2 & 0 & -0.5 \\ -1.9 & 0 & -0.9 & 0 & 0.3 & 0.5 & -0.2 \\ -0.2 & -1.1 & -1.1 & -1.2 & 0.1 & -0.6 & 0.3 \end{bmatrix}$$

In the existing literature, it is observed that the delay-free $H_\infty$ controller design for T-S fuzzy systems has been considered in [61, 63, 65, 67, 68]. Now, we evaluate the minimum allowable value of $\gamma$ obtained by using the methods proposed in [61, 63, 65, 67] and our paper. The comparison is given in Tables 2, where $\bar{d} = \bar{d}_i$, $i = 1,2,3$. It is seen from the Table that, for different values of the delay upper bounds, the minimum allowable values of $\gamma$ obtained by using our method are less than those obtained by using the methods in [61, 63, 65, 67] and our paper.

Table 2: Minimal Index $\gamma$ for $\mu_1, \mu_2 = 0.2$ and $\rho_1, \rho_2 = 100$

| Method | $d_M$ =0.5 | $d_M$ =0.6 | $d_M$ =0.8 | $d_M$ =1 |
|---|---|---|---|---|
| [61] | 0.38 | 0.43 | 0.83 | 2.22 |
| [63] | 0.36 | 0.39 | 0.51 | 0.79 |
| [65] | 0.33 | 0.35 | 0.43 | 0.53 |
| [67] | 0.31 | 0.33 | 0.37 | 0.44 |
| Corollary 1 | $10^{-3}$ | $10^{-2}$ | $10^{-2}$ | $10^{-1}$ |

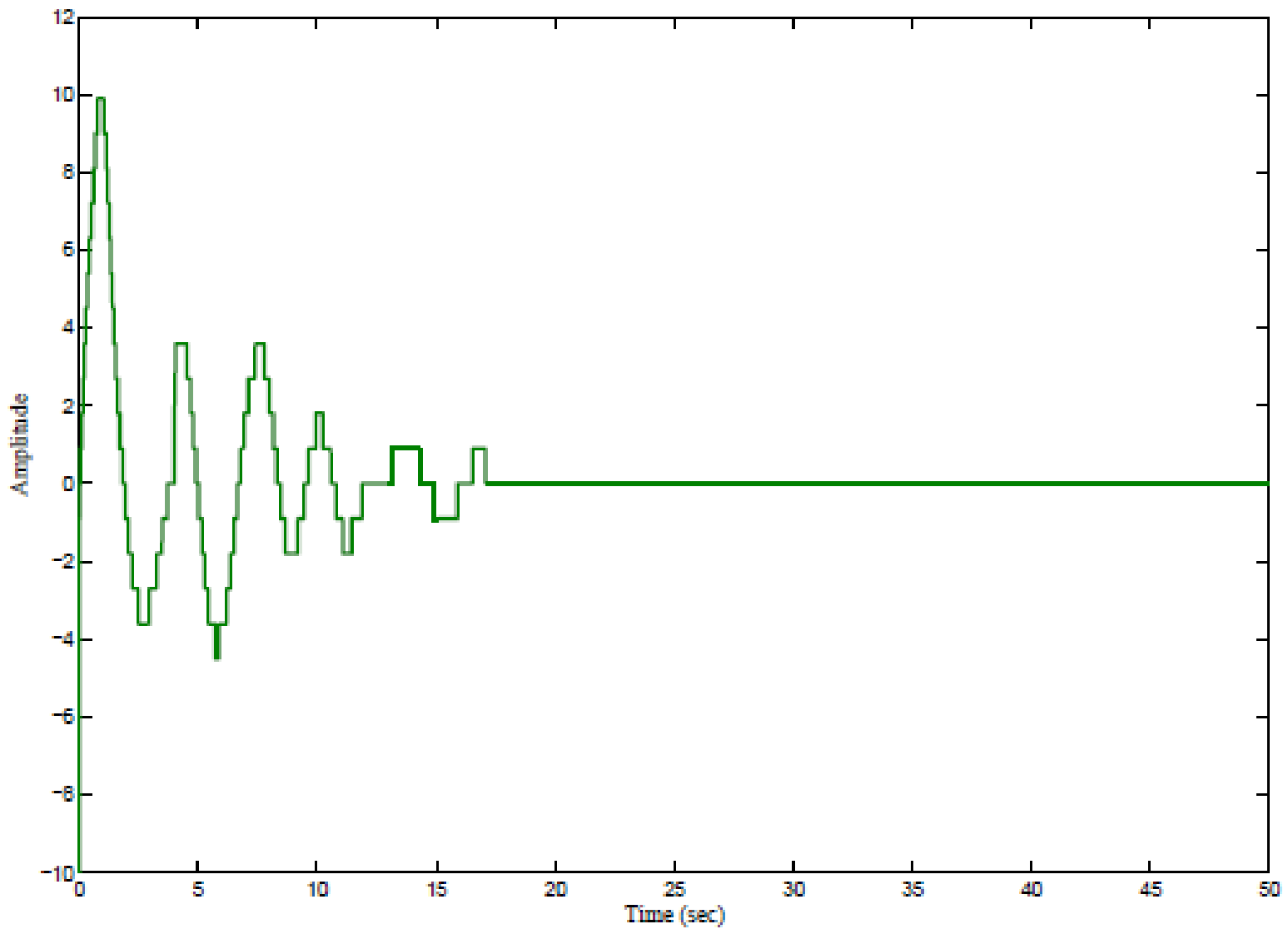

Figure 1: State Response of $x(t)$ for $H_\infty$ Control (Example 2.)

**Remark 2** *In [61, 63, 65], and [67], the authors list sets of* $\gamma_{min}$ *for various* $\delta$*. To elaborate our results with less conservativeness, we select their best results.*

## 4.2 Example 2

In this simulation example, we consider the characteristic of volt-ampere for the armature resistor of the nonlinear DC motor system (5) as (2), and the related parameters of the nonlinear DC motor system are given in Table 3.

Table 3: Nonlinear DC Motor System Parameters

| Parameter | $\mathbb{J}$ | $\mathbb{B}$ | $\mathbb{L}$ | $\mathbb{K}_a$ | $\mathbb{K}_b$ |
|---|---|---|---|---|---|
| Numerical values | 0.082 | 0.3 | 1000 | 0.576 | 0.612 |
| Unit | $kg.m^2$ | $N.m.s / rad$ | $mH$ | N.m/A | $V.s / rad$ |

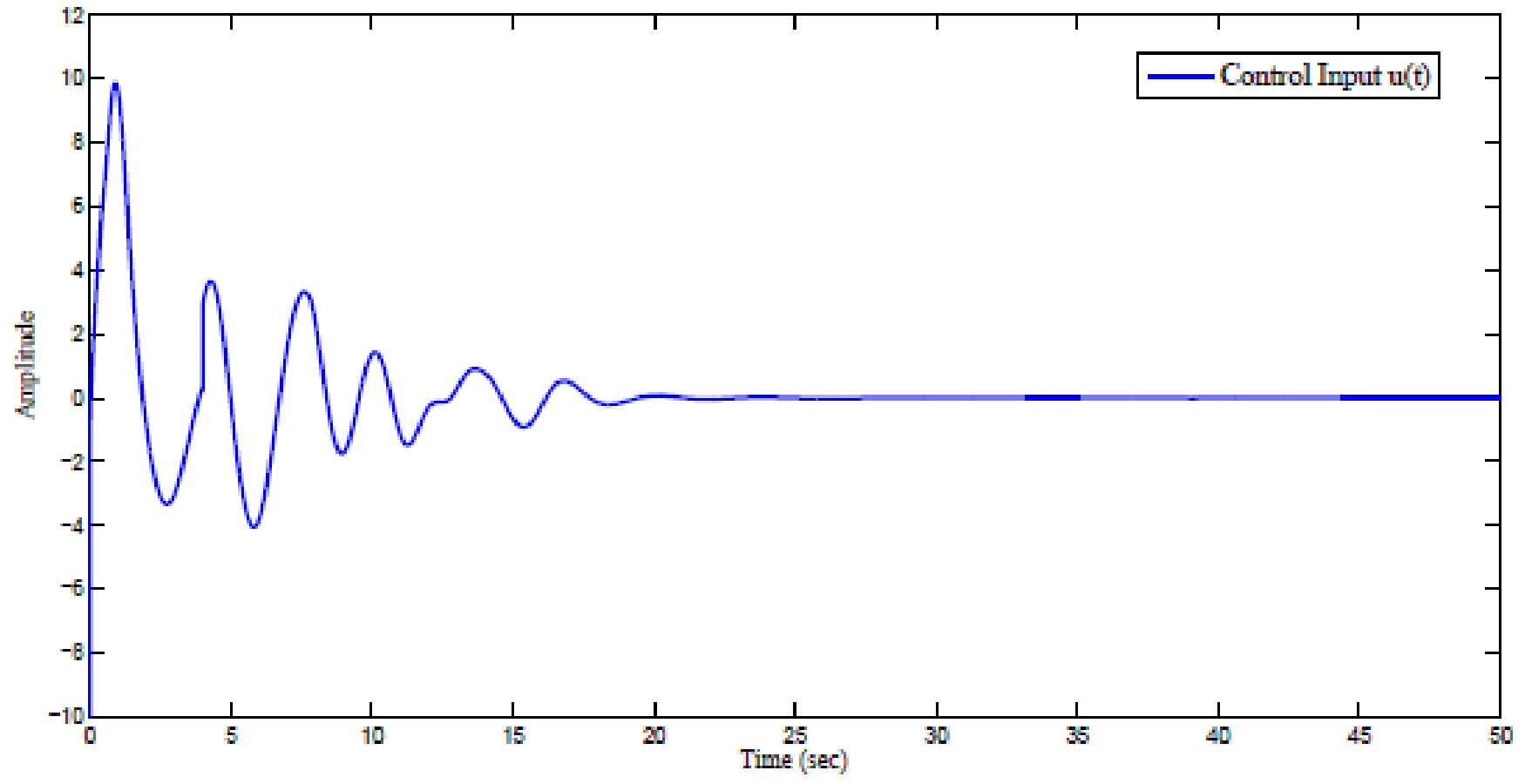

Figure 2: Control input $u(t)$ for $H_\infty$ Control (Example 2).

Let $x_1(t) = i(t)$ and $x_2(t) = \omega(t)$ be the state variable, then the system (5) with $u(t) = 0$ is signed by the following state equations:

$$\begin{cases} \mathbb{L}\dot{x}_1(t) = -u_R(t) - (\mathbb{K}_b + \nabla_b(t))x_2(t) \\ \mathbb{J}\dot{x}_2(t) = \mathbb{K}_a x_1(t) - Bx_2(t) + \omega(t) \end{cases} \tag{32}$$

According to given parameters in Table 3 and $\nabla_b(t) = 0.06b(t)(-1 \le b(t) \le 1)$, we have

$$\begin{cases} \dot{x}_1(t) = -4x_1(t) - x_1^3(t) - (0.612 + 0.06b(t))x_2(t) \\ \dot{x}_2(t) = \dfrac{0.576}{0.082}x_1(t) - \dfrac{0.3}{0.082}x_2(t) + \dfrac{1}{0.082}\omega(t) \end{cases} \quad (33)$$

The behavior of the system states $x_1(t)$ and $x_2(t)$ a in (33), with

$$w(t) = \begin{cases} 1, & 5 \le t \le 10 \\ -1, & 15 \le t \le 20 \\ 0, & else \end{cases}$$

and initial conditions $x_1(0) = 0.5$, $x_2(0) = -0.3$, is illustrated in Figure 4, which shows that the response converges bear off the stable equilibrium point $x_1(t) = x_2(t) = 0$, which yields that the system is stable.

Let us rewrite (33) as

$$\dot{x}(t) = \begin{bmatrix} -4 - x_1^2(t) & -(0.612 + 0.06b(t)) \\ \dfrac{0.576}{0.082} & -\dfrac{0.3}{0.082} \end{bmatrix} x(t) + \begin{bmatrix} 0 \\ -\dfrac{1}{0.082} \end{bmatrix} \omega(t) \quad (34)$$

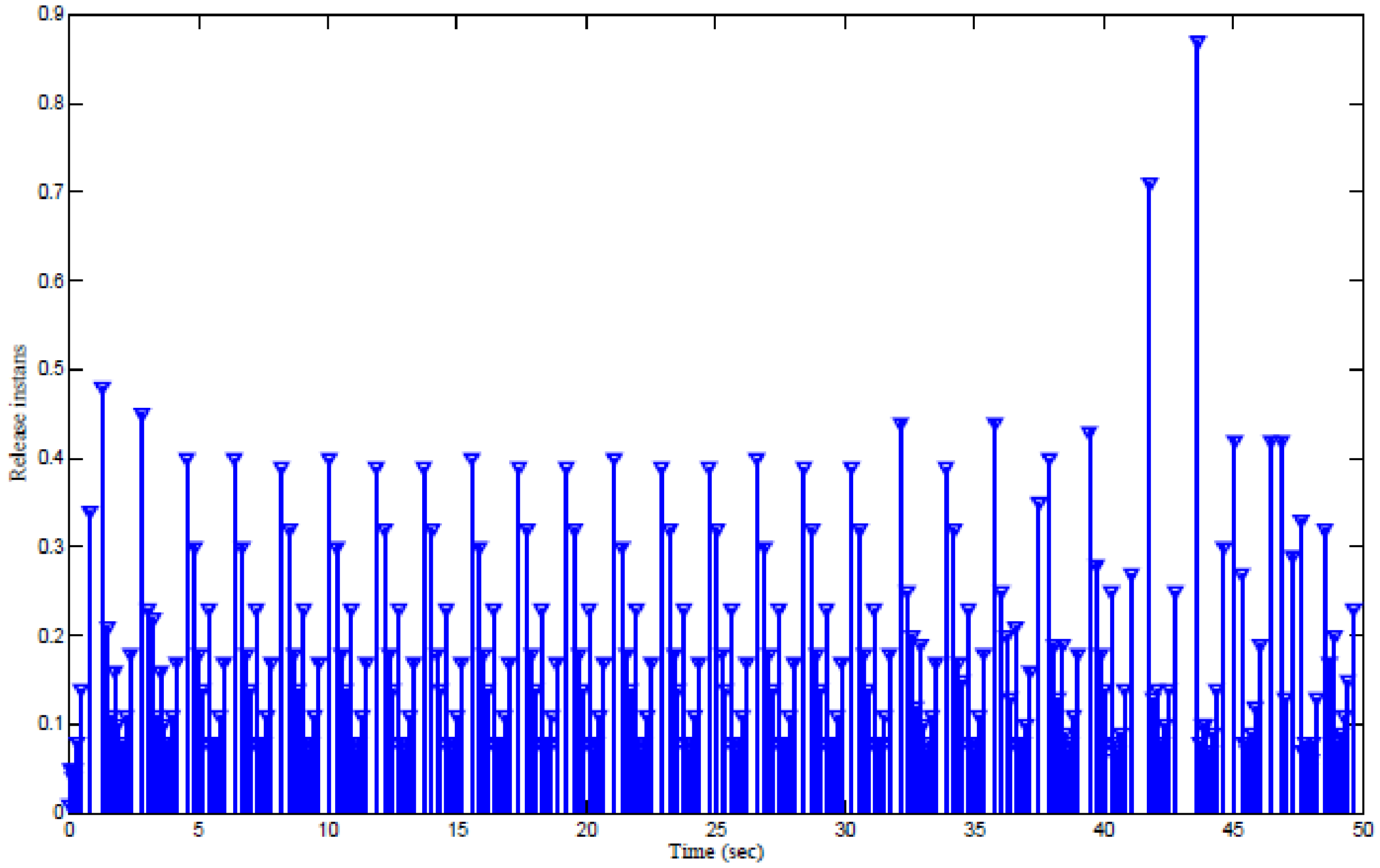

Figure 3: The release instants and intervals for $H_\infty$ Control (Example 2).

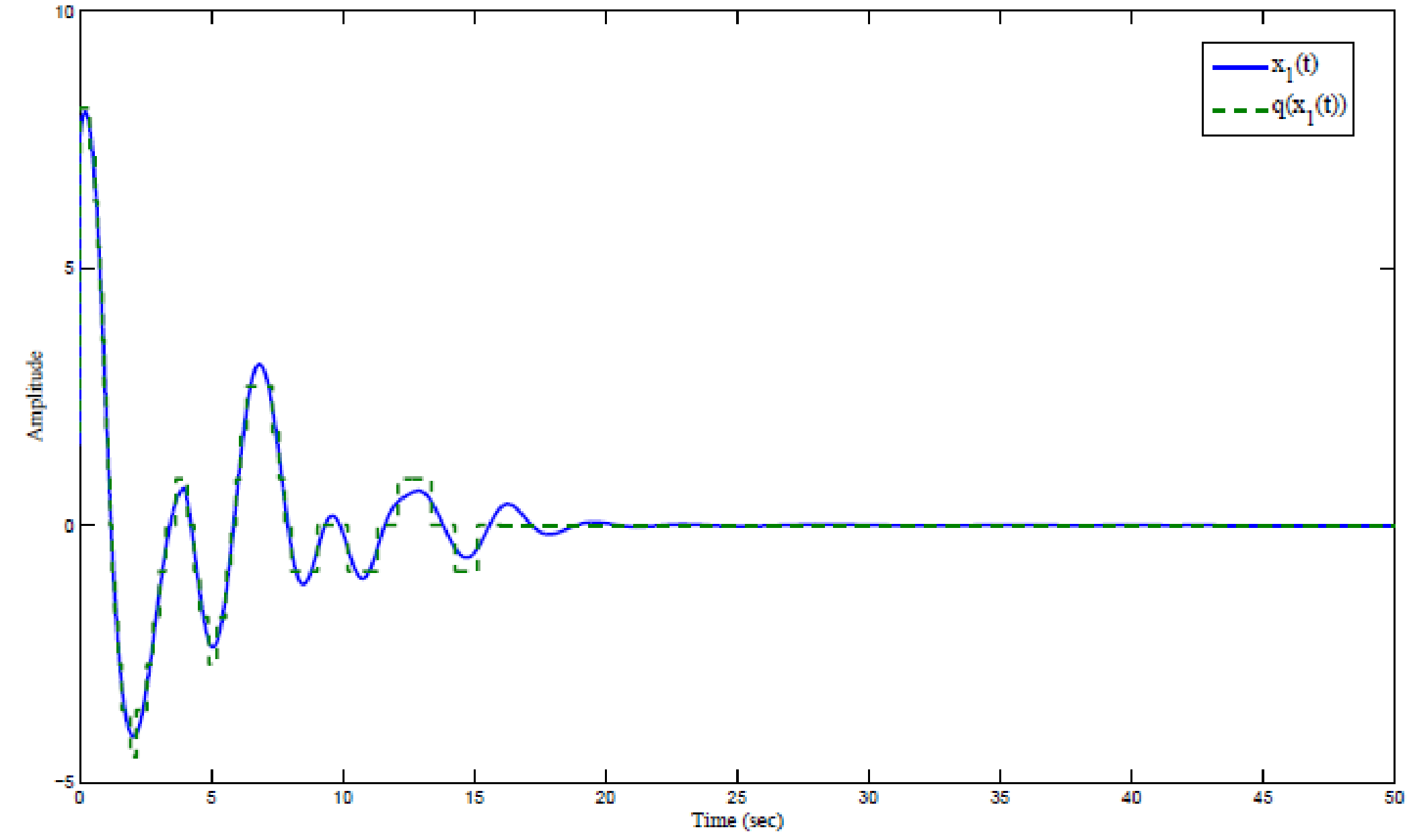

(a) State $x_1(t)$ with and without quantization (Example 2).

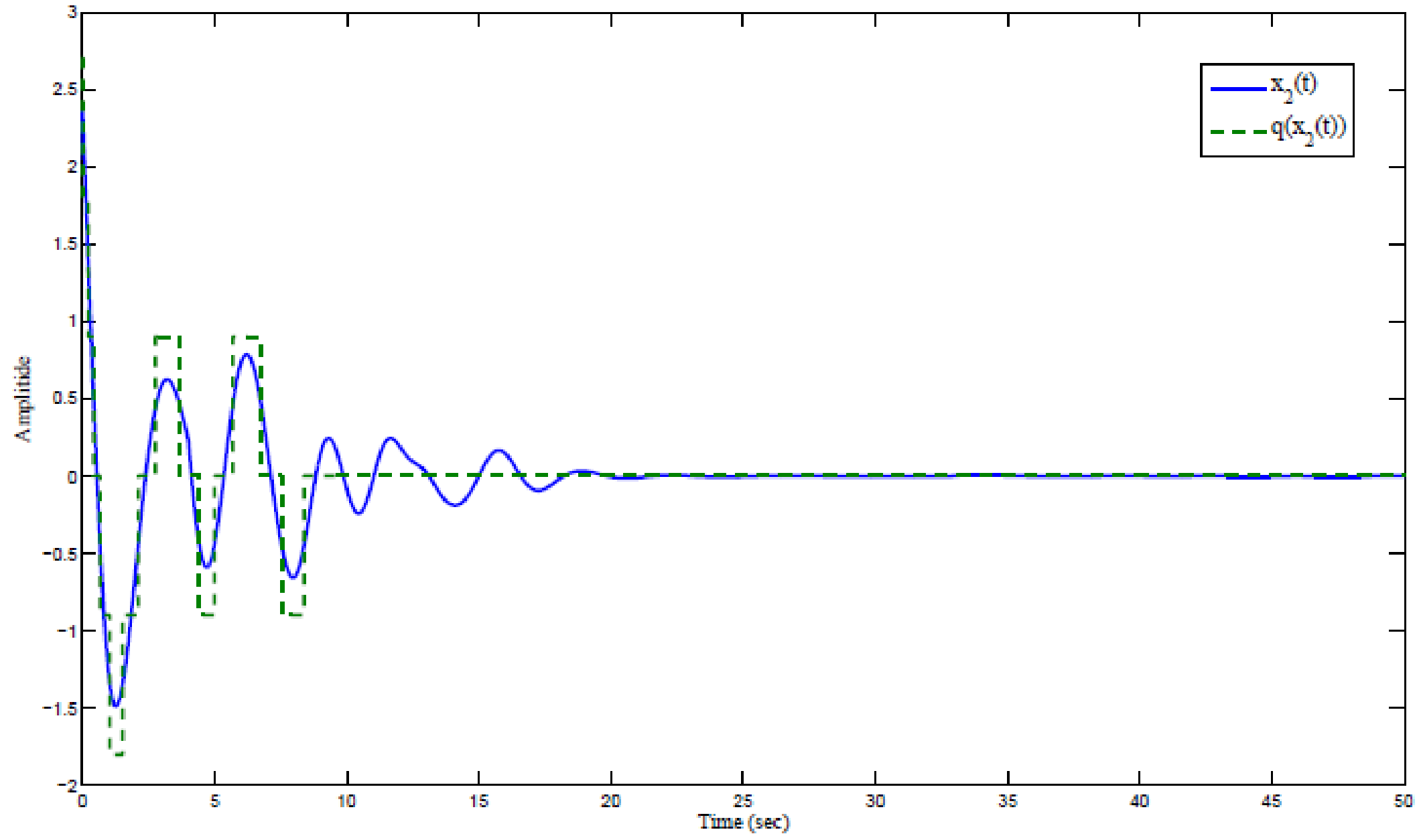

(b) State $x_2(t)$ with and without quantization (Example 2).

Figure 4: State with and without quantization (Example 2).

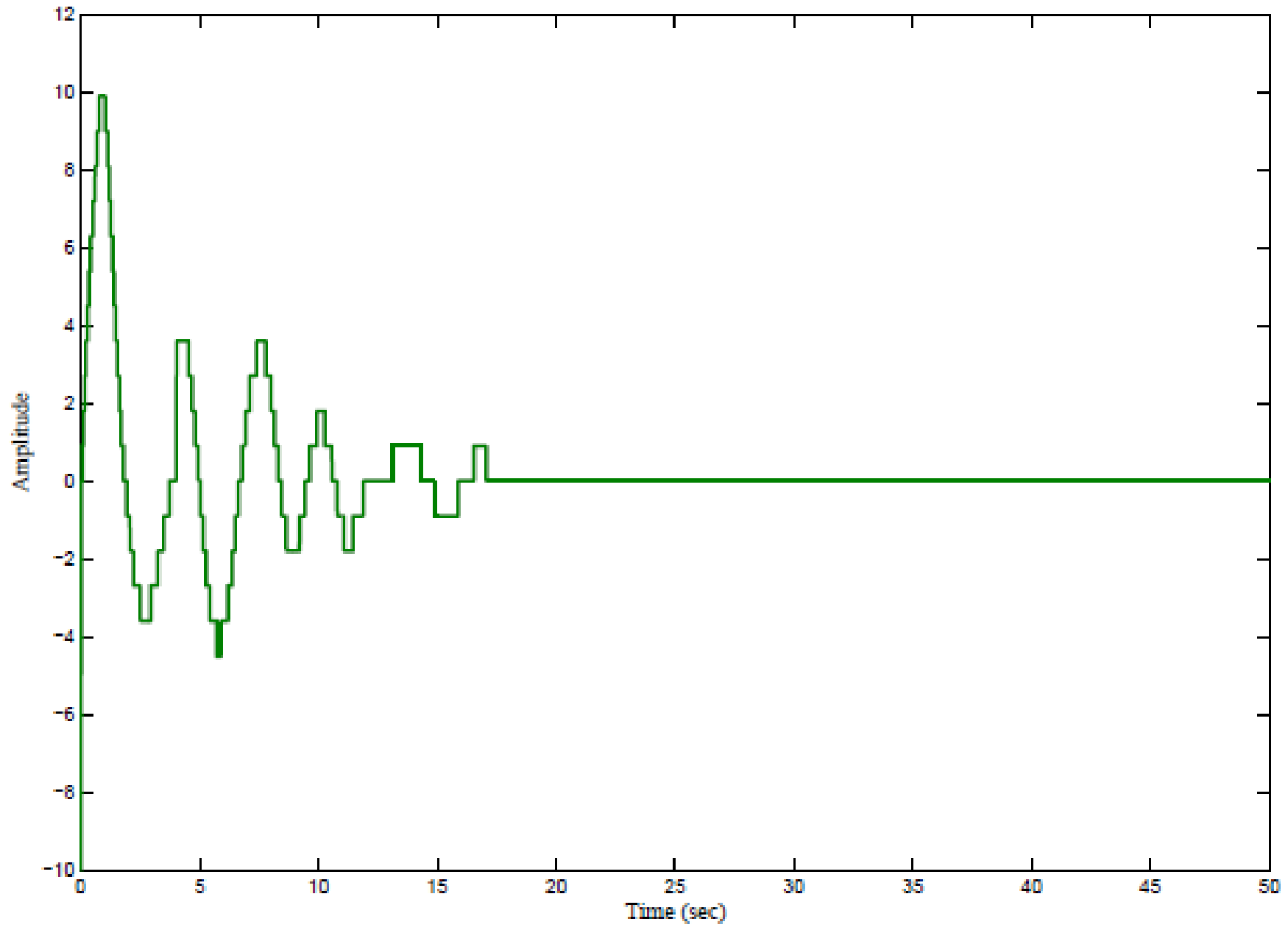

Figure 5: State $x_3(t)$ with and without quantization (Example 2).

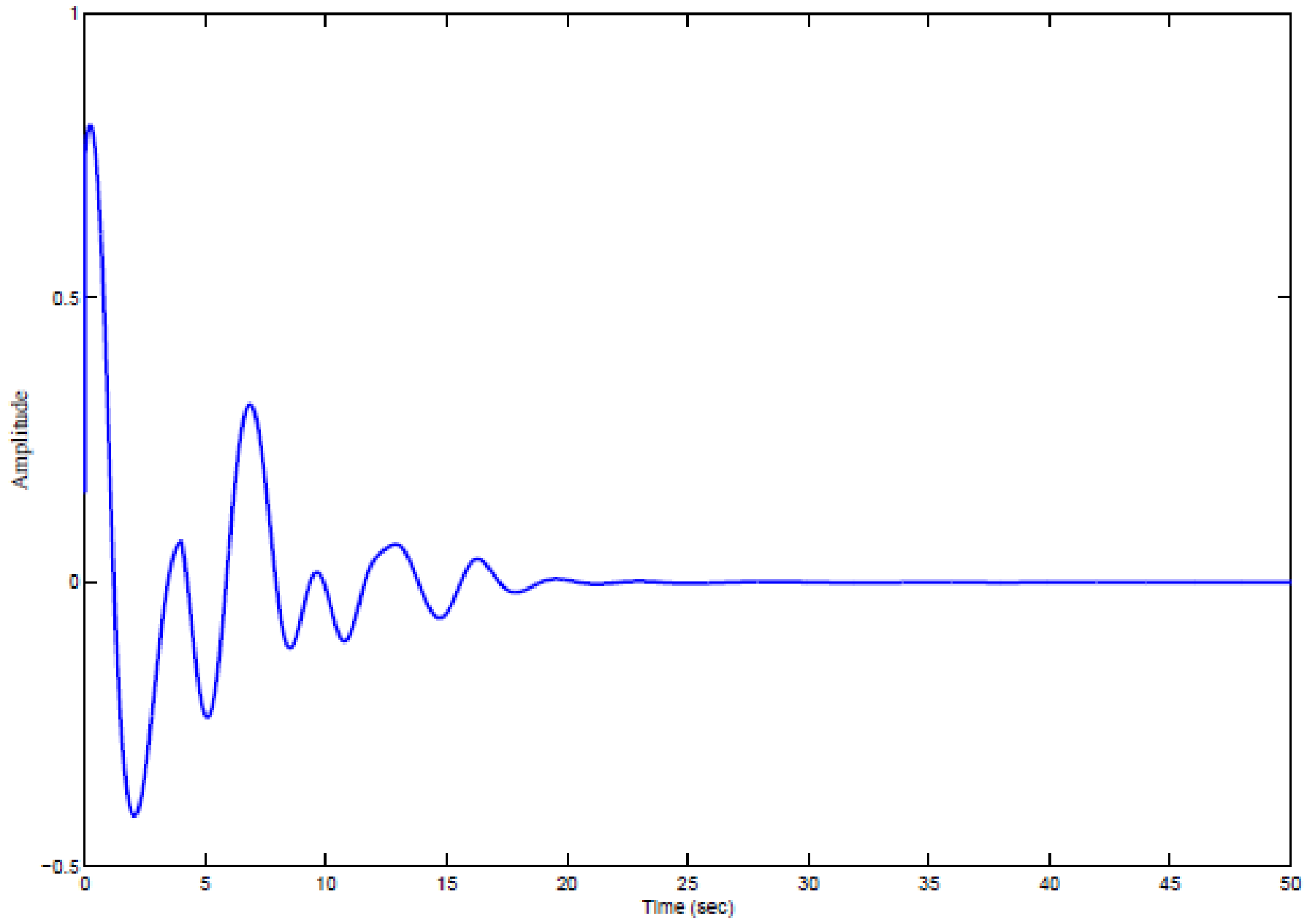

Figure 6: The states and quantized state responses for Truck-trailer system (Example 2)

For the nonlinear system (34), based on the sector nonlinearity approach, one can suppose that $-3 \le x_1(t) \le 3$, that is $-13 \le -4 - x_1^2(t) \le -4$. By employing the fuzzy modeling for the nonlinear system, the state equation (34) can be expressed by the T–S model with following fuzzy rules:

$$\dot{x}(t)=\sum_{i=1}^{r}\lambda_i(\zeta(t))(\mathbf{A}_i x(t)+B_i\omega(t)) \tag{35}$$

where $x(t)=[x_1(t) \quad x_1(t)]^T$ .

$$A_{1b}=\begin{bmatrix} -4 & -(0.612+0.06b(t)) \\ \frac{0.576}{0.082} & -\frac{0.3}{0.082} \end{bmatrix} \quad A_{2b}=\begin{bmatrix} -13 & -(0.612+0.06b(t)) \\ \frac{0.576}{0.082} & -\frac{0.3}{0.082} \end{bmatrix}$$

$$B_{1b}=B_{2b}=\begin{bmatrix} 0 \\ -\frac{1}{0.082} \end{bmatrix}$$

By applying the maximum and minimum values of $x_1(t)$ , $\lambda_1(\zeta_1(t))$ and $\lambda_2(\zeta_2(t))$ with $\zeta_1(t)=x_1(t)$ can be described as

$$x_1^2(t)=\lambda_1(\zeta_1(t))\times 9+\lambda_2(\zeta_1(t))\times 0 \tag{36}$$
$$\lambda_1(\zeta_1(t))+\lambda_2(\zeta_1(t))=1$$

Then, the membership function can be given by

$$\lambda_1(\zeta_1(t))=x_1^2(t)/9, \quad \lambda_2(\zeta_1(t))1-\lambda_1(\zeta_1(t)) \tag{37}$$

In the following, we will study the peak–to–peak filtering problem for the system (35). The T-S fuzzy system as (21) and the corresponding system parameter matrices are:

$$\mathbf{A}_1=\begin{bmatrix} 0.8 & -0.0306 \\ 0.3512 & 0.8171 \end{bmatrix}, \quad \mathbf{A}_2=\begin{bmatrix} 0.35 & -0.0306 \\ 0.3512 & 0.8171 \end{bmatrix}$$

$$B_1=B_2=\begin{bmatrix} 0 \\ -\frac{0.05}{0.082} \end{bmatrix}, \quad \mathbf{C}_1=[-3 \quad 0.5], \quad \mathbf{C}_2=[1.5 \quad 2]$$

$$\mathbf{E}_1=[0.2 \quad 1], \quad \mathbf{E}_2=[0.3 \quad 1]$$

with

$$\mathbf{A}_{d1}=\begin{bmatrix} -0.1 & 0.05 \\ -0.5 & -0.75 \end{bmatrix}, \quad \mathbf{A}_{d2}=\begin{bmatrix} -0.9 & 0 \\ -1.15 & -1.25 \end{bmatrix}, \quad \mathbf{E}_{d1}=[-0.1 \quad 0], \quad \mathbf{E}_{d2}=[0 \quad 0.2]$$

**Case-I** Nominal system. Considered the sensor failure with time varying delays $d_1(t)=0.01+0.002\sin(\frac{\pi t}{2}))$ , $d_2(t)=\frac{\bar{d}_2+\bar{d}_2\sin(2\mu_2 t/\bar{d}_2)}{2}$ and $\tau(t)=0.15+0.08sin((\pi t)/2)$ with $(\rho_1,\rho_2)=100$ and $(\mu_1,\mu_2,\beta,\delta_0)=(0.2,0.4,0.5,0.15)$ . Let $\gamma=1.5$ . Then, it is found that the LMIs (24)–(28) and (31) of Theorem 3 are feasible, and the feasible solutions to those LMIs are obtained as follows:

Table 4: Nonlinear DC Motor System Parameters

| Parameter | J | B | L | $K_a$ | $K_b$ |
|---|---|---|---|---|---|
| Numerical values | 0.082 | 0.3 | 1000 | 0.576 | 0.612 |
| Unit | $kg.m^2$ | $N.m.s/rad$ | $mH$ | N.m/A | $V.s/rad$ |

$$\begin{bmatrix} \mathbf{X}_1 & \mathbf{X}_2 \end{bmatrix} = \begin{bmatrix} 0.2439 & 0.4035 & 0.2671 & 0.4263 \\ 0.4035 & 1.5777 & 0.4263 & 1.9264 \end{bmatrix}$$

$$\begin{bmatrix} \mathbf{Y}_1 & \mathbf{Y}_2 \end{bmatrix} = \begin{bmatrix} 0.1467 & 0.1819 & 0.1156 & 0.1104 \\ 0.1819 & 0.5665 & 0.1104 & 0.3824 \end{bmatrix}$$

$$\begin{bmatrix} \breve{\mathbf{A}}_{f1} & \breve{\mathbf{A}}_{f2} \\ \breve{\mathbf{A}}_{\tau f1} & \breve{\mathbf{A}}_{\tau f2} \end{bmatrix} = \begin{bmatrix} 1.0271 & 0.4013 & 1.1210 & -0.4086 \\ -0.7848 & 1.9013 & 1.4657 & 1.4665 \\ -0.0097 & -0.0086 & 0.0229 & -0.0187 \\ -0.0625 & -0.0462 & 0.0440 & -0.0681 \end{bmatrix}$$

$$\begin{bmatrix} \breve{\mathbf{E}}_{f1} & \breve{\mathbf{E}}_{f2} \\ \breve{\mathbf{E}}_{\tau f1} & \breve{\mathbf{E}}_{\tau f2} \end{bmatrix} = \begin{bmatrix} 0.0380 & -0.1282 & -0.0980 & 0.4162 \\ -0.0129 & -0.0237 & 0.0092 & 0.0270 \end{bmatrix}$$

$$\begin{bmatrix} \breve{\mathbf{B}}_{f1} & \breve{\mathbf{B}}_{f2} \\ \breve{\mathbf{D}}_{f1} & \breve{\mathbf{D}}_{f2} \end{bmatrix} = \begin{bmatrix} -0.9023 & -1.1808 \\ 0.4023 & 0.4556 \\ 12.5211 & -12.1297 \end{bmatrix}$$

Now we are in stage to develop the $H_\infty$ controller design. Fuzzy filter parameters can be attained in the form of (30). By selecting in the initial condition of the filtering system $x_{f1}(0) = -0.5$ and $x_{f2}(0) = 0.3$, state responses of the filtering system with and without sensor fault are given in Figures 4 and 5, respectively. Furthermore, output behavior of filtering system is shown in Figure 6. To this end, route of estimation error and output of the system is depicted in Figure 7 .

**Case-II** Uncertain system. To construct the reliable robust $H_\infty$ controller design for singular T–S fuzzy system, we select the following parameters to compute the uncertain matrices:

$$M_1 = M_2 = \begin{bmatrix} -0.6 & 0.5 \end{bmatrix}^T, \quad N_{11} = N_{12} = \begin{bmatrix} 0 & 0.6 \end{bmatrix}, \quad N_{21} = N_{22} = \begin{bmatrix} 0.3 & 0.4 \end{bmatrix}$$

$$N_{31} = N_{32} = \begin{bmatrix} 0.4 & 0.5 \end{bmatrix}, \quad N_{41} = N_{42} = \begin{bmatrix} 0.5 & 0.6 \end{bmatrix}, \quad N_{51} = N_{52} = \begin{bmatrix} 0.15 & 0.25 \end{bmatrix}$$

$$N_{61} = N_{62} = \begin{bmatrix} 0.25 & 0.35 \end{bmatrix}, \quad N_{71} = N_{72} = \begin{bmatrix} 0.35 & 0.45 \end{bmatrix}, \quad N_{81} = N_{82} = \begin{bmatrix} 0.45 & 0.55 \end{bmatrix}$$

Then, it is found that the LMIs (24)–(28) and (31) of Theorem 3 are feasible, and the feasible solutions to those LMIs are obtained as follows:

$$\begin{bmatrix} \mathbf{X}_1 & \mathbf{X}_2 \end{bmatrix} = \begin{bmatrix} 6.3813 & -2.6465 & 4.6378 & -1.1721 \\ -2.6465 & 3.4899 & -1.1721 & 4.8614 \end{bmatrix}$$

$$\begin{bmatrix} \mathbf{Y}_1 & \mathbf{Y}_2 \end{bmatrix} = \begin{bmatrix} 2.4913 & -1.0776 & 2.7614 & -1.2595 \\ -1.0776 & 1.5518 & -1.2595 & 1.7496 \end{bmatrix}$$

$$\begin{bmatrix} \breve{\mathbf{A}}_{f1} & \breve{\mathbf{A}}_{f2} \\ \breve{\mathbf{A}}_{\tau f1} & \breve{\mathbf{A}}_{\tau f2} \end{bmatrix} = \begin{bmatrix} -0.5321 & 1.2327 & 0.0484 & -0.6257 \\ -0.8270 & -0.4452 & 0.1344 & 0.1952 \\ -0.0134 & 0.0072 & -0.0059 & 0.0050 \\ 0.0084 & -0.0058 & 0.0036 & -0.0002 \end{bmatrix}$$

$$\begin{bmatrix} \breve{\mathbf{E}}_{f1} & \breve{\mathbf{E}}_{f2} \\ \breve{\mathbf{E}}_{\tau f1} & \breve{\mathbf{E}}_{\tau f2} \end{bmatrix} = \begin{bmatrix} 0.0380 & -0.1282 & -0.0980 & 0.4162 \\ -0.0129 & -0.0237 & 0.0092 & 0.0270 \end{bmatrix}$$

$$\begin{bmatrix} \breve{\mathbf{B}}_{f1} & \breve{\mathbf{B}}_{f2} \\ \breve{\mathbf{D}}_{f1} & \breve{\mathbf{D}}_{f2} \end{bmatrix} = \begin{bmatrix} 0.7551 & 0.5013 \\ -0.5925 & -0.5844 \\ 6.231\times 10^{-4} & 4.371\times 10^{-3} \end{bmatrix}$$

To check the robustness of the singular fuzzy controller, we introduced the white gaussian noise. Purpose to incorporate this type of noise is that our desired controller received the signal over the communication channel, which is mostly equivalent this one. In order to exhibits the effective of the proposed controller, the simulation result is given in Fig. 6(a) and (b), in the presence of uncertainties. While on the other side, system output and estimation error approaches to zero, which shows that designed controller meet the system performance. In the existence of uncertainties of the network, which is especially important for fading channels.

## 5 Conclusions

In this article, we addressed the problem on reliable robust fuzzy control problem for random time-delays characterized by uncertain Takagi—Sugeno (T–S) fuzzy singular models. Notation of $H_\infty$ performance index has been used to guarantee the resulting filtering error system with new delay-dependent conditions, which is obtained based upon the new version of *fuzzy L–K functional*. It should be mentioned that the delay–dependent conditions have been derived for the existence of the desired delayed controllers , which were established in the form of general matrix inequalities and LMIs, respectively. Finally, a DC motor example has been demonstrated finally to present the effectiveness of the proposed reliable fuzzy singular controller. Furthermore, the singular fuzzy systems with stochastic effect governed by Levy noise or Poisson process is an open topic for the researchers. In future work, we will focus on improving the existing algorithms and searching for new methods [69, 70], and [71] to deal with the above probabilistic issues over the communication channel.